\font\bbb=msbm10 \font\bbs=msbm7                                   

\overfullrule=0pt

\input epsf.tex
\input pstricks  
\input color
\definecolor{Green}{rgb}{0.25,0.75,0.25}
\definecolor{review}{rgb}{0.25,0.75,0.25}
\definecolor{exercise}{rgb}{1,0,0}
\definecolor{lightgray}{rgb}{0.5,0.5,0.5}

\def\C{\hbox{\bbb C}}  
\def\N{\hbox{\bbb N}}  
  
\def\Z{\hbox{\bbb Z}} \def\sZ{\hbox{\bbs Z}}

\font\ssbf=cmssbx10

\def\IEEETIT{{\sl IEEE Trans.\ Inform.\ Theory\/}}

\def\JACM{{\sl J. ACM\/}}

\def\PRA{{\sl Phys.\ Rev.\ A\/}}

\def\PRL{{\sl Phys.\ Rev.\ Lett.}}
\def\PRSLA{{\sl Proc.\ Roy.\ Soc.\ Lond.\ A\/}}

\def\QIP{{\sl Quantum Inform.\ Processing\/}}

\def\SIAMJC{{\sl SIAM J. Comput.}}

\def\TC{{\sl Theory of Computing\/}}
\def\TCS{{\sl Theor.\ Comput.\ Science\/}}

\def\dajm{\hbox{D. A. Meyer}}

\def\dj{\hbox{\dajm\ and J. Pommersheim}}

\def\grover{\hbox{L. K. Grover}}

\def\simon{\hbox{D. R. Simon}}

\def\hfb{\hfil\break}

\catcode`@=11
\newskip\ttglue

   \font\ninerm=cmr9    \font\eightrm=cmr8   \font\sixrm=cmr6
  \font\ninebf=cmbx9   \font\eightbf=cmbx8  \font\sixbf=cmbx6
  \font\nineit=cmti9   \font\eightit=cmti8  
  \font\ninesl=cmsl9   \font\eightsl=cmsl8  
  \font\ninemi=cmmi9   \font\eightmi=cmmi8  \font\sixmi=cmmi6

\font\bigten=cmr10 scaled\magstep2 

\def\ninepoint{\def\rm{\fam0\ninerm}%
  \textfont0=\ninerm \scriptfont0=\sixrm
  \textfont1=\ninemi \scriptfont1=\sixmi
  \textfont\itfam=\nineit  \def\it{\fam\itfam\nineit}%
  \textfont\slfam=\ninesl  \def\sl{\fam\slfam\ninesl}%
  \textfont\bffam=\ninebf  \scriptfont\bffam=\sixbf
    \def\bf{\fam\bffam\ninebf}%
  \tt \ttglue=.5em plus.25em minus.15em
  \normalbaselineskip=11pt
  \setbox\strutbox=\hbox{\vrule height8pt depth3pt width0pt}%
  \normalbaselines\rm}

\def\eightpoint{\def\rm{\fam0\eightrm}%
  \textfont0=\eightrm \scriptfont0=\sixrm
  \textfont1=\eightmi \scriptfont1=\sixmi
  \textfont\itfam=\eightit  \def\it{\fam\itfam\eightit}%
  \textfont\slfam=\eightsl  \def\sl{\fam\slfam\eightsl}%
  \textfont\bffam=\eightbf  \scriptfont\bffam=\sixbf
    \def\bf{\fam\bffam\eightbf}%
  \tt \ttglue=.5em plus.25em minus.15em
  \normalbaselineskip=9pt
  \setbox\strutbox=\hbox{\vrule height7pt depth2pt width0pt}%
  \normalbaselines\rm}

\def\sfootnote#1{\edef\@sf{\spacefactor\the\spacefactor}#1\@sf
      \insert\footins\bgroup\eightpoint
      \interlinepenalty100 \let\par=\endgraf
        \leftskip=0pt \rightskip=0pt
        \splittopskip=10pt plus 1pt minus 1pt \floatingpenalty=20000
        \parskip=0pt\smallskip\item{#1}\bgroup\strut\aftergroup\@foot\let\next}
\skip\footins=12pt plus 2pt minus 2pt
\dimen\footins=30pc

\def\ie{{\it i.e.}}
\def\eg{{\it e.g.}}

\def\etal{{\it et al.}}

\def\Lemma{L{\eightpoint EMMA}}

\def\Theorem{T{\eightpoint HEOREM}}

\def\Proposition{P{\eightpoint ROPOSITION}}
\def\endproof{\vrule height6pt width4pt depth2pt}

\def\and{{\eightpoint AND}}

\def\PARITY{P{\eightpoint ARITY}}
\def\SUM{S{\eightpoint UM}}

\def\Deutsch{1}
\def\CEMM{2}
\def\BBCMdW{3}
\def\Simon{4}
\def\Kuperberg{5}
\def\AMR{6}
\def\BCvD{7}
\def\Grover{8}
\def\SKW{9}
\def\AaronsonAmbainis{10}
\def\BHMT{11}
\def\MeyerPommersheim{12}
\def\vanDam{13}
\def\HunzikerMeyer{14}
\def\vanDamSeroussi{15}
\def\Shakeel{16}
\def\YKL{17}
\def\MeyerPommersheimbounds{18}

\magnification=1200
\input epsf.tex
\parskip=0pt\parindent=0pt

\centerline{\bf\bigten Multi-query quantum sums}
\bigskip
\centerline{David A. Meyer$^1$ and James Pommersheim$^{1,2}$}
\bigskip 
\centerline{\sl {}$^1$Department of Mathematics}
\centerline{\sl University of California/San Diego,
                La Jolla, CA 92093-0112}
\smallskip
\centerline{\sl {}$^2$Department of Mathematics}
\centerline{\sl Reed College, Portland, OR 97203}
\smallskip
\centerline{{\tt dmeyer@math.ucsd.edu},
            {\tt jamie@reed.edu}}
\bigskip\bigskip
{\bf Abstract}.  P{\eightpoint ARITY} is the problem of determining 
the parity of a string $f$ of $n$ bits given access to an oracle that 
responds to a query $x\in\{0,1,\ldots,n-1\}$ with the $x^{\rm th}$ bit
of the string, $f(x)$.  Classically, $n$ queries are required to 
succeed with probability greater than $1/2$ (assuming equal prior 
probabilities for all length $n$ bitstrings), but only 
$\lceil n/2\rceil$ quantum queries suffice to determine the parity 
with probability $1$.  We consider a generalization to strings $f$ of
$n$ elements of $\Z_k$ and the problem of determining $\sum f(x)$.  By
constructing an explicit algorithm, we show that $n-r$ ($n\ge r\in\N$)
{\sl entangled\/} quantum queries suffice to compute the sum correctly 
with worst case probability $\min\{\lfloor n/r\rfloor/k,1\}$.  This 
quantum algorithm utilizes the $n-r$ queries sequentially and 
adaptively, like Grover's algorithm, but in a different way that is 
not amplitude amplification.

\medskip
\parskip=10pt
\noindent{\bf 1.  Introduction}

\PARITY\ is the oracle (or black-box) problem of determining the 
parity of an $n$-bit string by querying positions in the string.  
Since even a single unqueried bit can change the parity, $n$ classical
queries are required to solve this problem with probability $1$, 
assuming all $n$-bit strings are possible.

When $n = 2$, this is Deutsch's problem [\Deutsch], for which a single
quantum query, used properly, suffices [\CEMM].  Beals \etal\ show 
that in general $\lceil n/2\rceil$ quantum queries suffice by applying 
the solution to Deutsch's problem to the bits in pairs [\BBCMdW].  In
their algorithm the quantum queries are {\sl independent\/} of one 
another---they can be asked in parallel since none depends on the 
responses of the oracle to the others---and they are also 
{\sl incoherent}---after each query is processed, the state is 
measured and the resulting information (the parity of a pair of the 
bits) is combined classically at the end of the algorithm.

This same independence of multiple queries is a feature of existing 
multi-query quantum algorithms for abelian [\Simon] and non-abelian 
(\eg, [\Kuperberg,\AMR,\BCvD]) hidden subgroup problems, which range 
from incoherent [\Simon] through partially [\Kuperberg,\AMR] to 
completely [\BCvD] coherent.  Grover's quantum search algorithm 
[\Grover], and quantum (random walk) search algorithms on graphs 
[\SKW,\AaronsonAmbainis] more generally, however, utilize coherent 
sequences of {\sl adapted\/} queries---the quantum state is modified by 
each oracle response before it is returned to the oracle for the next 
query, so the queries are not independent.  These algorithms all use 
{\sl amplitude amplification\/} [\BHMT] to adapt their sequential 
queries.

But amplitude amplification, which identifies an element in the 
preimage of $1$ for some bit-valued function, does not apply to 
\PARITY, nor to its generalization:

{\global\setbox3=\hbox{\SUM.\enspace}
\parindent=\wd3\narrower\item{\SUM.}  
Let $f:\Z_n\to\Z_k$, where $f$ is accessed {\it via\/} an oracle that 
responds with $f(x)$ when queried about $x\in\Z_n$.  Find 
$\sum_{x\in\sZ_n}f(x)$ (modulo $k$).\par}

As they are for \PARITY, $n-1$ classical queries are {\sl useless\/}
for \SUM\ when $f$ is chosen uniformly at random, \ie, the $1/k$ prior 
probability of each possible sum is unchanged after the oracle 
responds to the queries [\MeyerPommersheim].  Our Uselessness Theorem:
if $2q$ classical queries are useless, then $q$ quantum queries are 
useless [\MeyerPommersheim], implies that $\lfloor (n-1)/2\rfloor$ 
quantum queries are therefore useless for \SUM.  This raises the 
question of how well we can do using more than a useless number of 
queries; to answer it we construct an $n-r$ quantum query algorithm 
that computes the sum correctly with worst case probability 
$\min\{\lfloor n/r\rfloor/k,1\}$, for each $1\le r\in\N$, and that 
returns a result that is within $\lfloor kr/2n\rfloor$ of the sum 
with probability at least $4/\pi^2$.  This quantum algorithm utilizes 
the $n-r$ queries sequentially and adaptively, like quantum search 
algorithms, but in a different way that is not amplitude 
amplification.

We motivate the development of our algorithm in the next section by 
considering the simplest new instances of \SUM, computing the sum of 
2 or 3 trits.  In \S3 we state and prove two basic lemmas and combine 
them to construct the general algorithm in \S4.  We conclude in \S5 by 
recalling the result of van Dam that strings of $n$ bits can be 
identified with high probability using $n/2 + O(\sqrt{n})$ queries, 
and hence any function of them can be computed with at least the same 
probability [\vanDam].  We generalize this result to $k > 2$ and show
that, unsurprisingly---since it is designed to do more than just 
compute the sum of the string values, it gives success probabilities 
less than those of our algorithm.

\medskip
{\bf 2.  Sums of trits}

The simplest generalization of Deutsch's problem is to add two trits 
rather than two bits, \ie, the $n = 2$ and $k = 3$ version of \SUM.
As with Deutsch's problem, if all possible functions $f:\Z_2\to\Z_3$ 
are equally likely, a single classical query is useless---the prior
probabilities of $1/3$ for each value of $\sum f(x)$ are unchanged 
after a single query---while two classical queries suffice to 
determine the sum with probability 1.  Thus the goal of a quantum 
algorithm for this problem should be to determine the sum with a 
single quantum query with probability greater than $1/3$.

\Proposition\ 1.  {\sl Using a single quantum query the sum of two 
trits can be determined with worst case probability $2/3$.}

Before giving the proof we recall some standard notation:  We will 
work in the Hilbert space $\C^n\otimes\C^k$, with computational basis 
$\bigl\{|x\rangle|y\rangle\,\big|\, x\in\Z_n, y\in\Z_k\bigr\}$.  The 
shift operator acts by $X : |z\rangle \mapsto |z+1\rangle$ and the 
oracle acts by 
${\cal O}_{\!f} : |x\rangle|y\rangle \mapsto |x\rangle|y+f(x)\rangle
= |x\rangle X^{f(x)}|y\rangle$.  Finally, $\omega = e^{2\pi i/k}$, and 
the Fourier transform on $\C^k$ acts by
$$
{\cal F}:|y\rangle
 \mapsto 
{1\over\sqrt{k}}\sum_{\ell=0}^{k-1}\omega^{\ell y}|\ell\rangle 
 =: 
|\omega^{-y}\rangle,                                          \eqno(1)
$$
since $X|\omega^{-y}\rangle = \omega^{-y}|\omega^{-y}\rangle$.  These
``Fourier'' (or ``character'') basis states will be used to implement
the widely useful generalization to dimensions greater than 2 
[\HunzikerMeyer,\vanDamSeroussi,\Shakeel] of the ``phase 
kickback trick'' [\CEMM].  

To use these states in a quantum algorithm for the two trit problem we 
might expect simply to query the oracle with a state of the form
$$
{1\over\sqrt{2}}\bigl(|0\rangle + |1\rangle\bigr)
 \otimes
|\phi\rangle,
$$
where $\phi = \omega^{-y}$ for some $y\in\{1,2\}\subset\Z_3$, as if we 
were solving Deutsch's problem.  Notice, however, that the relative 
phase of the two components in the state returned by the
oracle would be $\phi^{f(0)-f(1)}$, so it would not encode the sum 
$f(0) + f(1)$, unlike the two bit case in which 
$-f(1) \equiv f(1)$~(mod 2).  A different query state is required:

{\sl Proof\/} (of Proposition 1).  It suffices to exhibit a single 
query algorithm that succeeds with probability 2/3.

\global\setbox2=\hbox{{\bf 0}.\enspace}
\parindent=2\wd2\parskip=5pt
\item{{\ssbf 0}.} 
Initialize to the state 
$$
{1\over\sqrt{2}}\bigl(|1\rangle|\omega^1\rangle + 
                      |0\rangle|\omega^{-1}\rangle 
                \bigr).                                       \eqno(2)
$$
                               
\item{{\ssbf 1}.}
Call the oracle ${\cal O}_{\!f}$ to obtain the state
$$
{1\over\sqrt{2}}\bigl(\omega^{f(1)}|1\rangle|\omega^1\rangle + 
                      \omega^{-f(0)}|0\rangle|\omega^{-1}\rangle
                \bigr).
$$
Notice that the relative phase of the two terms is 
$\omega^{f(0)+f(1)}$.  We could argue at this point that there is a 
POVM that identifies which of the three possible states we have with
probability 2/3 [\YKL], but as a simple sequence of unitary 
transformations avoids the necessity for anything beyond a complete 
von Neumann measurement in the computational basis, we describe it 
explicitly in the following steps.

\item{{\ssbf 2}.}  
Act by $X\otimes I$ to obtain the state
$$
{1\over\sqrt{2}}\bigl(\omega^{f(1)}|0\rangle|\omega^1\rangle + 
                      \omega^{-f(0)}|1\rangle|\omega^{-1}\rangle
                \bigr).
$$

\item{{\ssbf 3}.}
Act by $K$ to obtain the state
$$
{1\over\sqrt{2}}|0\rangle\bigl(\omega^{f(1)}|\omega^1\rangle + 
                               \omega^{-f(0)}|\omega^0\rangle 
                         \bigr),                              \eqno(3)
$$
where $K$ acts on $\C^n\otimes\C^k$ by
$$
K:|x\rangle|\omega^y\rangle 
 =
\cases{|0\rangle|\omega^0\rangle      
                                    & if $x = n-1$ and $y = k-1$;  \cr
       |n-1\rangle|\omega^{k-1}\rangle 
                                         & if $x = 0$ and $y = 0$; \cr
       |x\rangle|\omega^y\rangle
                                                    & otherwise.   \cr
      }
$$
Note that while $K$ is a complicated unitary operation, it is 
independent of $f$, \ie, it does not call the oracle.

\parindent=0pt
The $\C^3$ tensor factor in the final state (3) can be rewritten as:
$$ 
\eqalignno{
&{1\over\sqrt{2}}\bigl(\omega^{f(1)}|\omega^1\rangle + 
                      \omega^{-f(0)}|\omega^0\rangle 
                \bigr)                                             \cr
&\qquad\qquad\qquad\qquad=  
\omega^{-f(0)}{1\over\sqrt{2}}
  \bigl(\omega^{\Sigma f}|\omega^1\rangle + 
        |\omega^0\rangle 
  \bigr)                                                       &(4)\cr
&\qquad\qquad\qquad\qquad=
\omega^{-f(0)}{1\over\sqrt{6}}
\Bigl(\bigl(1+\omega^{\Sigma f}\bigr)|0\rangle +
      \bigl(1+\omega^{\Sigma f-1}\bigr)|1\rangle +
      \bigl(1+\omega^{\Sigma f-2}\bigr)|2\rangle
\Bigr),                                                            \cr
}
$$
using the definition (1), so now measurement of the $\C^3$ tensor 
factor will return $\sum f(x)$ with probability $2/3$. \hfill\endproof

\parskip=10pt
To obtain this probability our initial query (2) was an 
{\sl entangled\/} state, rather than the usual tensor product state;
this is the first innovation in the algorithm up to which we are 
building.  The next step is to consider adding $n = 3$ trits.  In this
case two classical queries are useless, so one quantum query is 
useless [\MeyerPommersheim], and we must consider algorithms with two 
coherent quantum queries.

\Proposition\ 2.  {\sl Two quantum queries suffice to solve {\rm\SUM}\ 
with probability $1$ when $n = k = 3$.}

{\sl Proof}.  It suffices to exhibit a two query algorithm that 
succeeds with probability 1.

\global\setbox2=\hbox{{\bf 0}.\enspace}
\parindent=2\wd2\parskip=5pt
\item{{\ssbf 0}.} 
Initialize to the entangled state 
$$
{1\over\sqrt{3}}\bigl(|1\rangle|\omega^1\rangle + 
                      |0\rangle|\omega^{-1}\rangle +
                      |0\rangle|\omega^{-2}\rangle
                \bigr).
$$
                               
\item{{\ssbf 1}.}
Call the oracle ${\cal O}_{\!f}$ to obtain the state
$$ 
{1\over\sqrt{3}}\bigl(\omega^{f(1)}|1\rangle|\omega^1\rangle + 
                      \omega^{-f(0)}|0\rangle|\omega^{-1}\rangle +
                      \omega^{-2f(0)}|0\rangle|\omega^{-2}\rangle
                \bigr).
$$

\item{{\ssbf 2}.}  
Act by $X\otimes I$ to obtain the state
$$ 
{1\over\sqrt{3}}\bigl(\omega^{f(1)}|2\rangle|\omega^1\rangle + 
                      \omega^{-f(0)}|1\rangle|\omega^{-1}\rangle +
                      \omega^{-2f(0)}|1\rangle|\omega^{-2}\rangle
                \bigr).
$$

\item{{\ssbf 3}.}  
Act by $J_1$ to obtain the state
$$ 
{1\over\sqrt{3}}\bigl(\omega^{f(1)}|2\rangle|\omega^2\rangle + 
                      \omega^{-f(0)}|2\rangle|\omega^1\rangle +
                      \omega^{-2f(0)}|1\rangle|\omega^{-1}\rangle
                \bigr),                                       \eqno(5)
$$
where $J_r$ acts on $\C^n\otimes\C^k$ by
$$
J_r:|x\rangle|\omega^y\rangle 
 =
\cases{|x\rangle|\omega^0\rangle                    & if $y = 0$;  \cr
       |x+r\rangle|\omega^1\rangle 
                                                    & if $y = -1$; \cr
       |x\rangle|\omega^{y+1}\rangle
                                                    & otherwise.   \cr
      }
$$
Note that like $K$, while $J_r$ is a complicated unitary operation, it 
is independent of $f$, \ie, it does not call the oracle.

\item{{\ssbf 4}.}  
Call the oracle ${\cal O}_{\!f}$ a second time to obtain the state
$$ 
{1\over\sqrt{3}}\bigl(\omega^{f(1)+2f(2)}|2\rangle|\omega^2\rangle + 
                      \omega^{-f(0)+f(2)}|2\rangle|\omega^1\rangle +
                      \omega^{-2f(0)-f(1)}|1\rangle|\omega^{-1}\rangle
                \bigr).
$$

\item{{\ssbf 5}.}  
Act by $X\otimes I$ again to obtain the state
$$ 
{1\over\sqrt{3}}\bigl(\omega^{f(1)+2f(2)}|0\rangle|\omega^2\rangle + 
                      \omega^{-f(0)+f(2)}|0\rangle|\omega^1\rangle +
                      \omega^{-2f(0)-f(1)}|2\rangle|\omega^{-1}\rangle
                \bigr).
$$

\item{{\ssbf 6}.}
Act by $K$ to obtain the state
$$ 
{1\over\sqrt{3}}|0\rangle\bigl(\omega^{f(1)+2f(2)}|\omega^2\rangle + 
                               \omega^{-f(0)+f(2)}|\omega^1\rangle +
                               \omega^{-2f(0)-f(1)}|\omega^0\rangle
                         \bigr).                              \eqno(6)
$$

\parindent=0pt
The $\C^3$ tensor factor in the final state (6) can be rewritten as:
$$ 
\eqalignno{
&{1\over\sqrt{3}}\bigl(\omega^{f(1)+2f(2)}|\omega^2\rangle + 
                       \omega^{-f(0)+f(2)}|\omega^1\rangle +
                       \omega^{-2f(0)-f(1)}|\omega^0\rangle
                 \bigr)                                            \cr
&\qquad\qquad\qquad\qquad=
{1\over\sqrt{3}}\omega^{f(0)+2f(1)}
  \bigl(\omega^{-\Sigma f}|\omega^2\rangle + 
        \omega^{-2\Sigma f}|\omega^1\rangle +
        \omega^{-3\Sigma f}|\omega^0\rangle
  \bigr)                                                       &(7)\cr
&\qquad\qquad\qquad\qquad=
\omega^{f(0)+2f(1)}|\Sigma\,f\rangle,                              \cr
}
$$
using the definition (1), so now measurement of this tensor factor 
will return $\sum f(x)$ with probability $1$.          \hfill\endproof

\parskip=10pt
The key piece of algebra is that the phases of the terms in (6), each
a linear combination of two values of $f$, are also linear 
combinations of all {\sl three\/} values of $f$, with a coefficient of 
$0$ in front of the third value:  $(0,1,2)\cdot{\bf f}$, 
$(-1,0,1)\cdot{\bf f} = (2,0,1)\cdot{\bf f}$, and 
$(-2,-1,0)\cdot{\bf f} = (1,2,0)\cdot{\bf f}$, where 
${\bf f} = \bigl(f(0),f(1),f(2)\bigr)$.  Written this way it is clear 
that the coefficient vectors are successive cyclic shifts $\sigma$ of 
$(0,1,2)$, so if we factor out the last phase factor the other two 
become:
$$
\eqalign{
(0,1,2)\cdot{\bf f} - \sigma^2(0,1,2)\cdot{\bf f} 
 &= (2,2,2)\cdot{\bf f} = -\sum f                                  \cr 
\sigma(0,1,2)\cdot{\bf f} - \sigma^2(0,1,2)\cdot{\bf f} 
 &= \sigma(1,1,1)\cdot{\bf f} = -2\sum f,                          \cr
}
$$
the phases of the first two terms in (7).

This algorithm is optimal since it uses only one more than the useless 
number of quantum queries.  Notice that its two coherent quantum 
queries are {\sl sequential\/} rather than parallel, and that the 
second query is {\sl adapted\/} in the sense that the state (5) that 
is passed to the oracle as the second query depends on the response of 
the oracle to the first query, unitarily transformed by 
$J(X\otimes I)$.  This adaptation differs from amplitude amplification
[\BHMT] and is the second innovation in our quantum summation 
algorithm.

\medskip
{\bf 3.  Two basic lemmas}

To generalize the quantum algorithms given in the previous section for 
summing trits, it is convenient first to state two basic lemmas.

\Lemma\ 3.  {\sl For $A\in\Z_k$ and $k\ge s\in\N$, let 
$$
|A_s\rangle 
 = 
{1\over\sqrt{s}}
\sum_{\ell=1}^s \omega^{-\ell A}|\omega^{s-\ell}\rangle
\in\C^k.                                                      \eqno(8)
$$
Measurement of $|A_s\rangle$ in the computational basis returns 
$|y\rangle$, $y\in\Z_k$, with probability
$$
\bigl|\langle y|A_s\rangle\bigr|^2
 =
{1\over sk}\Bigl({\sin\pi s(y-A)/k\over\sin\pi(y-A)/k}\Bigr)^2,
                                                              \eqno(9)
$$
defined to be a continuous function of $y-A$.  The probability 
$\bigl|\langle y|A_s\rangle\bigr|^2$ takes its maximum value, $s/k$, 
at $y = A$, and the probability that the measurement is within 
$\pm\lfloor k/2s\rfloor$ of $A$ is at least $4/\pi^2$.}

{\sl Proof}.  This is an elementary (and familiar from phase 
estimation; see, \eg, [\CEMM]) calculation using the definition (1):
$$
\langle y|A_s\rangle 
 =
{1\over\sqrt{sk}}\sum_{\ell=1}^s\omega^{-\ell A - (s-\ell)y}
 =
{1\over\sqrt{sk}}\,\omega^{-sy}\sum_{\ell=1}^s\omega^{\ell(y-A)}
 =
{1\over\sqrt{sk}}\,\omega^{(1-s)y-A}{1-\omega^{s(y-A)}
                                     \over
                                     1-\omega^{y-A}
                                    }.
$$
Taking the norm squared of this expression gives (9), which by 
continuity takes the value $s/k$ when $y = A$.  That this is the 
maximum follows from the fact that in this case all the terms in the
sum above are 1.

Writing $d = y-A$, $|d|\le k/2s$ implies 
$|\sin\pi sd/k| \ge |\pi sd/k|/(\pi/2) = 2s|d|/k$, since the argument
of $\sin$ has absolute value no more than $\pi/2$.  Also, 
$|\sin\pi d/k| \le |\pi d/k|$.  Using these bounds in (9) gives
$$
\bigl|\langle A+d|A_s\rangle\bigr|^2 
 \ge
{1\over sk}\Bigl({2s|d|/k\over |\pi d/k|}\Bigr)^2 
 = 
{4\over\pi^2}{s\over k},
$$
so
$$
\sum_{|d|\le k/2s}{4\over\pi^2}{s\over k} 
 \ge 
\Bigl\lceil{k\over s}\Bigr\rceil\cdot{4\over\pi^2}{s\over k} 
 \ge
{4\over\pi^2}.                                          \eqno\endproof
$$

When $k = 3 = s$ and $A = \sum f(x)$, the state (8) is equal to the 
$\C^3$ tensor factor in the final state of the algorithm in 
Proposition~2, up to an overall phase.  Similarly, in the algorithm of
Proposition~1, if rather than factoring out the phase $\omega^{-f(0)}$ 
in (4), we factor out $\omega^{f(0)+2f(1)}$, we obtain (8) with 
$k = 3$, $s = 2$, and $A = \sum f(x)$.  The success probabilities of 
$1$ and $2/3$ in these two algorithms are the values $s/k$ given by 
Lemma~3.

When $A = \sum_{x\in\sZ_n} f(x)$, each component of the state (8) 
depends on all of the $n$ values of $f$.  The next lemma says that, up
to an overall phase, this state is equivalent to one in which each 
component depends on fewer than $n$ values of $f$.

\Lemma\ 4.  {\sl Let $1\le r\in\N$, let $r|n$, and let $s = n/r$.
Then   
$$
\eqalign{
&\omega^{\Sigma_{m=1}^s 
                          m[f((m-1)r)+\cdots+f(mr-1)]
                    }
 {1\over\sqrt{s}}\sum_{\ell=1}^s
                 \omega^{-\ell\Sigma f}|\omega^{s-\ell}\rangle \cr
&\qquad\qquad\qquad\qquad\qquad\qquad=
{1\over\sqrt{s}}
\sum_{\ell=1}^s
\omega^{\Sigma_{m=1}^s 
                         (m-\ell)[f((m-1)r)+\cdots+f(mr-1)]
                   }|\omega^{s-\ell}\rangle. \cr}
$$
}

For each value of $\ell$, namely for each component, in the sum on the 
right hand side of this equation, there is a term in the sum in the
exponent which vanishes because $m = \ell$.  Since each of these terms
depends on $r$ values of $f$, each component depends on 
$sr - r = n - r$ values of $f$.  In the algorithm of Proposition~1, 
$n = 2$ and $r = 1$ so $s = n/r = 2$, and each component of the final 
state (3) depends on $n - r = 2 - 1 = 1$ value of $f$.  In the 
algorithm of Proposition~2, $n = 3$ and $r = 1$ so $s = n/r = 3$, and 
each component of the final state (6) depends on $n - r = 3 - 1 = 2$ 
values of $f$.

\medskip
{\bf 4.  The general S{\eightbf UM} problem}

Summing two and three trits are special cases of the general \SUM\ 
problem that motivate the two innovations in our general algorithm.
Propositions~1 and 2 are special cases of the following theorem.

\Theorem\ 5.  {\sl Let $f:\Z_n\to\Z_k$.  Using $n-r$ quantum queries 
the sum $\sum_{x\in\sZ_k}f(x)$ can be computed correctly with worst
case probability $\min\{\lfloor n/r\rfloor/k,1\}$, for each 
$n\ge r\in\N$.  Furthermore, the same algorithm outputs a result 
within $\lfloor kr/2n\rfloor$ of the correct sum with probability at 
least $4/\pi^2$.}

{\sl Proof}.  First consider the case $r|n$ and let $s = n/r\in\N$.  
If $s \le k$ and we can construct the state 
$$
{1\over\sqrt{s}}\sum_{\ell=1}^s
                \omega^{-\ell\Sigma f}|\omega^{s-\ell}\rangle,
$$
then by Lemma~3 we can find $\sum f(x)$ with probability 
$s/k = n/(rk)$, and even when the output is wrong, it is likely to be
close---within $\lfloor rk/2n\rfloor$ with probability at least 
$4/\pi^2$.  By Lemma~4 we need only construct the state
$$
{1\over\sqrt{s}}
\sum_{\ell=1}^s
 \omega^{\Sigma_{m=1}^s (m-\ell)[f((m-1)r)+\cdots+f(mr-1)]}
 |\omega^{s-\ell}\rangle,
$$
in which each component depends on $n-r$ values of $f$.  The 
following algorithm does so, using $n-r$ quantum queries:

\global\setbox2=\hbox{{\bf 0}.\enspace}
\parindent=2\wd2\parskip=5pt
\item{{\ssbf 0}.} 
Initialize to the entangled state 
$$
{1\over\sqrt{s}}\bigl(|r\rangle|\omega^1\rangle +
                      |0\rangle|\omega^{-1}\rangle +
                        \cdots +
                      |0\rangle|\omega^{-(s-1)}\rangle
                \bigr).
$$

\item{{\ssbf 1}.} 
Apply 
$K\bigl(\bigl((X\otimes I){\cal O}_{\!f}\bigr)^r\bigr)
  \bigl(J_r\bigl((X\otimes I){\cal O}_{\!f}\bigr)^r\bigr)^{s-2}$ to
obtain the state
$$
{1\over\sqrt{s}}
|0\rangle
\sum_{\ell=1}^s
\omega^{\Sigma_{m=1}^s (m-\ell)[f((m-1)r)+\cdots+f(mr-1)]}
        |\omega^{s-\ell}\rangle.
$$

\item{{\ssbf 2}.}
Measure the $\C^k$ tensor factor in the computational basis.

\parindent=0pt
Notice that when $n = k$ and $r = 1$, \ie, using $k - 1$ quantum 
queries, this algorithm returns $\sum f(x)$ with probability $1$.

\parindent=0pt\parskip=10pt
If $s > k$, or equivalently, if $r < n/k$, then $n = uk + v$ with 
$u \ge r$ and $0 \le v < k$, so we can use $k-1$ queries in this 
algorithm applied to each block of length $k$, using a total of 
$uk - u = n - v - u$ queries, leaving $v+u-r \ge v$ queries to 
identify the last $v$ values of $f$.  Thus when $s > k$, we can find 
$\sum f(x)$ with probability $1$.

Second, and similarly, if $r\!\!\not\kern-0.4pt|\;n$, let 
$s = \lfloor n/r\rfloor$.  Then $n = rs + w$ with $0 < w < r$.  Using 
the algorithm applied to the first $n-w$ values of $f$, we can compute
$$
\sum_{x=0}^{rs-1}f(x)\hbox{, with probability min}%
\bigl\{1,\lfloor n/r\rfloor/k\bigr\},
$$
using $n-w-r$ queries, leaving $w$ queries to identify the last $w$ 
values of $f$.

Thus in all cases, this algorithm uses $n-r$ quantum queries to return 
$\sum f(x)$ with probability 
min$\bigl\{1,\lfloor n/r\rfloor/k\bigr\}$, and a value within 
$\lfloor kr/2n\rfloor$ of the sum with probability at least $4/\pi^2$.   
                                                       \hfill\endproof
                                                       
We believe this algorithm is optimal, but we have only proved it to be
so for $r = n-1$, \ie, a single query [\MeyerPommersheimbounds].                                                      

\medskip
{\bf 5.  Conclusion}

Since the number of queries $n-r \ge 0$, the success probability of 
our algorithm is always at least $1/k$, as it should be.  Furthermore, 
since $\lfloor n/r\rfloor = 1$ until $r \le n/2$, fewer than $n/2$ 
quantum queries in this algorithm are useless, as they must be 
according to the Uselessness Theorem [\MeyerPommersheim].  When 
$k = 2$, Theorem~5 says that for $r \le n/2$ the success probability 
is 1, as we know from the solution to \PARITY\ [\BBCMdW].

For $k > 2$ we know of no algorithms to which to compare ours.  Van
Dam's quantum algorithm for obtaining all the information about a 
function $\Z_n\to\Z_2$ with high probability using $n/2+O(\sqrt{n})$
queries [\vanDam], however, can be generalized to functions 
$f:\Z_n\to\Z_k$:

\Theorem\ 6.  {\sl Let $f:\Z_n\to\Z_k$.  There is a quantum algorithm 
using $q$ queries that correctly identifies the function with worst
case probability 
$$
p_q = {1\over k^n}\sum_{j=0}^q {n\choose j}(k-1)^j.          \eqno(10)
$$
}

The cumulative distribution function (10) for this binomial 
probability distribution is greater than 0.95 (almost 0.98) provided 
$q > n(k-1)/k + 2\sqrt{n(k-1)}/k$, namely the mean plus two standard
deviations.  Thus with this many queries we can determine the oracle
correctly with probability more than 0.95, and thus compute the sum of
its values correctly.  More precisely, using this algorithm with $q$
queries, we can compute $\sum f(x)$ with probability less than
$p_q + (1-p_q)/k$ (obtained by bounding the probability of computing 
the sum correctly by $1/k$ when the algorithm fails to output the 
correct $f$).  Figure~1 plots this upper bound on the success 
probability as a function of the number of queries, along with the 
success probability of the algorithm of Theorem~5.

The success probability of the algorithm of Theorem~5 is greater than 
or equal to that of the generalized van Dam algorithm of Theorem~6, 
for any number of queries, an unsurprising result since the latter is 
using those queries to try to determine the whole function, not just 
its sum.  To succeed with probability greater than a constant, the 
former requires a fraction of $n$ approaching $1$ like $1/k$ quantum 
queries, while the latter requires this many {\sl plus\/} 
$O(\sqrt{n})$.

\vskip2\baselineskip
\hbox to 6.5truein{\hfill%
\hbox{\epsfxsize=5truein\epsfbox{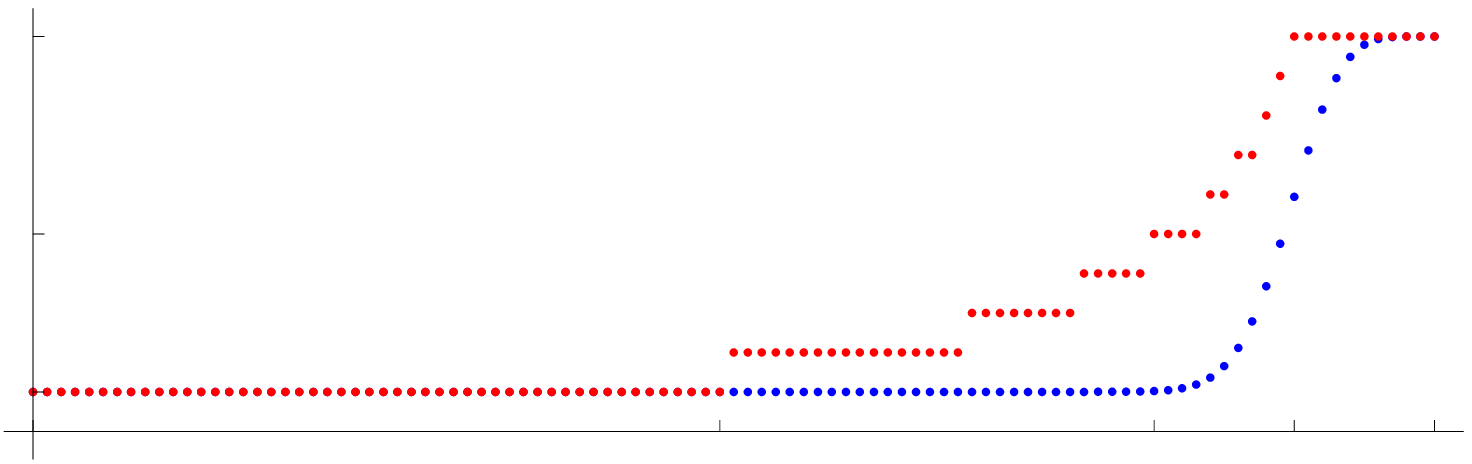}}\hfill}
\vskip-10\baselineskip
\hskip0.7truein {\ninepoint prob.}\hfb
\vskip-1.85\baselineskip
\hskip0.72truein {\eightpoint 1}\hfb
\vskip0.55\baselineskip
\hskip0.56truein {\eightpoint $1/2$}\hfb
\vskip-0.1\baselineskip
\hskip0.56truein {\eightpoint $1/k$}\hskip4.35truein\raise4pt\hbox{\eightpoint $n-\lfloor n/k\rfloor$}
\vskip-1.2\baselineskip
\hskip5.85truein {\ninepoint $q$}\hfb
\vskip-1.91\baselineskip
\hskip2.82truein {\eightpoint $\lfloor{n-1\over2}\rfloor$
\hskip0.7truein $\lceil n(1-{2\over k})\rceil$
\hskip0.85truein\raise4pt\hbox{\eightpoint $n$}}\hfb

\centerline{\ninepoint{\ninebf Fig.~1}.  Success probabilities of the
algorithms from Theorems~5 (steps) and 6 (smooth).} 

Finally, to succeed only {\sl approximately} (\ie, within 
$\epsilon k$) with probability greater than a constant ($4/\pi^2$; the 
sum of probabilities calculated for the central peak of (9) in
Lemma~3), the algorithm of Theorem~5 requires $n(1-\epsilon)$ quantum 
queries, independent of $k$.

\medskip
\noindent{\bf Acknowledgement}

This work has been partially supported by the Defense Advanced 
Research Projects Agency as part of the Quantum Entanglement Science
and Technology program under grant N66001-09-1-2025.

\medskip
\global\setbox1=\hbox{[00]\enspace}
\parindent=\wd1
\noindent{\bf References}
\vskip10pt

\parskip=0pt

\item{[\Deutsch]}
D. Deutsch,
``Quantum theory, the Church-Turing principle and the universal 
  quantum computer'',
\PRSLA\ {\bf 400} (1985) 97--117.

\item{[\CEMM]}
R. Cleve, A. Ekert, C. Macchiavello and M. Mosca,
``Quantum algorithms revisited'',
\PRSLA\ {\bf 454} (1998) 339--354.

\item{[\BBCMdW]}
R. Beals, H. Buhrman, R. Cleve, M. Mosca and R. de Wolf,
``Quantum lower bounds by polynomials'',
\JACM\ {\bf 48} (2001) 778--797.

\item{[\Simon]}
\simon,
``On the power of quantum computation'',
in S. Goldwasser, ed.,
{\sl Proceedings of the 35th Annual Symposium on Foundations of 
     Computer Science}, Santa Fe, NM, 20--22 November 1994
(Los Alamitos, CA:  IEEE 1994) 116--123;
%
\SIAMJC\ {\bf  26} (1997) 1474--1483.

\item{[\Kuperberg]}
G. Kuperberg, 
``A subexponential-time quantum algorithm for the dihedral hidden 
  subgroup problem'',
{\tt quant-ph/0302112}; 
\SIAMJC\ {\bf 35} (2005) 170--188.

\item{[\AMR]}
G. Alagic, C. Moore and A. Russell, 
``Quantum algorithms for Simon's problem over general groups'',
{\tt quant-ph/0603251}; 
in
{\sl Proceedings of the 18th Annual ACM-SIAM Symposium on Discrete 
     Algorithms}, New Orleans, LA, 7--9 January 2007
(New York \& Philadelphia:  ACM \& SIAM 2007) 1217--1224.

\item{[\BCvD]}
D. Bacon, A. M. Childs and W. van Dam, 
``From optimal measurement to efficient quantum algorithms for the 
  hidden subgroup problem over semidirect product groups'',
{\tt quant-ph/0504083}; 
in
{\sl Proceedings of the 46th Annual Symposium on Foundations of 
     Computer Science}, Pittsburgh, PA, 22--25 October 2005
(Los Alamitos, CA:  IEEE 2005) 469--478.

\item{[\Grover]}
\grover,
``A fast quantum mechanical algorithm for database search'',
in 
{\sl Proceedings of the Twenty-Eighth Annual Symposium on the Theory 
     of Computing}, Philadelphia, PA, 22--24 May 1996 
(New York:  ACM 1996) 212--219;\hfb
\grover, 
``Quantum mechanics helps in searching for a needle in a haystack'', 
{\tt quant-ph/9706033};
\PRL\ {\bf 79} (1997) 325--328.

\item{[\SKW]}
N. Shenvi, J. Kempe and K. B. Whaley,
``Quantum random walk search algorithm'',
{\tt quant-ph/0210064};
\PRA\ {\bf 67} (2003) 052307/1--11.

\item{[\AaronsonAmbainis]}
S. Aaronson and A. Ambainis,
``Quantum search of spatial regions'',
{\tt quant-ph/ 0303041};
in
{\sl Proceedings of the 44th Annual Symposium on Foundations of 
     Computer Science}, Cambridge, MA, 11--14 October 2003
(Los Alamitos, CA:  IEEE 2003) 200--209; 
\TC\ {\bf 1} (2005) 47--79.

\item{[\BHMT]}
G. Brassard, P. H{\o}yer, M. Mosca and A. Tapp,
``Quantum amplitude amplification and estimation'',
{\tt quant-ph/0005055};
in
S. J. Lomonaco, Jr.\ and H. E. Brandt, eds.,
{\sl Quantum Computation and Information},
{\sl Contemporary Mathematics\/} {\bf 305}
(Providence, RI:  AMS 2002) 53--74.

\item{[\MeyerPommersheim]}
\dj,
``On the uselessness of quantum queries'',
{\tt arXiv: 1004.1434 [quant-ph]};
to appear in \TCS.

\item{[\vanDam]}
W. van Dam,
``Quantum oracle interrogation: getting all information for almost 
  half the price'',
{\tt quant-ph/9805006};
in
{\sl Proceedings of the 39th Annual Symposium on Foundations of 
     Computer Science}, Palo Alto, CA, 8--11 November 1998
(Los Alamitos, CA:  IEEE 1998) 362--367.

\item{[\HunzikerMeyer]}
M. Hunziker and \dajm,
``Quantum algorithms for highly structured search problems'',
\QIP\ {\bf 1} (2002) 145--154.

\item{[\vanDamSeroussi]}
W. van Dam and G. Seroussi,
``Efficient quantum algorithms for estimating Gauss sums'',
{\tt quant-ph/0207131}.

\item{[\Shakeel]}
A. Shakeel,
``An improved query for the hidden subgroup problem'',
{\tt arXiv:1101.1053 [quant-ph]}.

\item{[\YKL]}
H. P. Yuen, R. S. Kennedy and M. Lax,
``Optimum testing of multiple hypotheses in quantum detection 
  theory'',
\IEEETIT\ {\bf IT-21} (1975) 125--134.

\item{[\MeyerPommersheimbounds]}
\dj, in preparation.
 
\bye